\documentclass[sigconf,authorversion,nonacm]{acmart}
\AtBeginDocument{%
  }

\newcommand{\wei}[1]{\textcolor{brown}{#1}}

\usepackage{multirow}

\begin{document}

\title{Benchmarking StreamNTT with a Verilog-to-Routing Toolchain}

\author{Wei He}
\affiliation{%
  \institution{Rochester Institute of Technology}
  \city{Rochester}
  \state{NY}
  \country{USA}
}
\email{wh9297@rit.edu}

\author{\mbox{Young-kyu Choi, Hyunwoo Park}}
\affiliation{%
  \institution{Inha University}
  \city{Incheon}
  \country{South Korea}
}
\email{ykc@inha.ac.kr, hwpark2000@gmail.com}

\author{Sunwoong Kim}
\affiliation{%
  \institution{Rochester Institute of Technology}
  \city{Rochester}
  \state{NY}
  \country{USA}
}
\email{sskeme@rit.edu}    


\begin{abstract}
As post-quantum cryptography algorithms move toward large-scale data center deployment, hardware acceleration of their computational bottleneck, which is the number theoretic transform (NTT), has gained increasing attention. 
StreamNTT, a high-level synthesis- and field-programmable gate array-based accelerator, achieves state-of-the-art throughput through various optimization techniques.
However, its reliance on a commercial tool and a device makes direct comparisons difficult for researchers without access.
We address this by building StreamNTT on an open-source Verilog-to-Routing toolchain, which achieves similar digital signal processing and multiplier usage.
Significant differences in internal memory utilization indicate that further memory-level optimization is needed to approach commercial tool performance.
\end{abstract}

\maketitle

\section{Introduction}
%
%
As post-quantum cryptography (PQC) standards, such as CRYSTALS-Kyber and CRYSTALS-Dilithium, move toward large-scale deployment in data centers~\cite{sosnowski2023performance}, the number theoretic transform (NTT), which is the dominant kernel for polynomial multiplications in lattice-based PQC, faces growing demand for higher throughput.

%
NTT is based on integer arithmetic and involves numerous butterfly operations. 
Each butterfly operation involves a computationally expensive modular multiplication that can be implemented using a parallel and pipelined architecture \cite{kim2019fpga}. 
In addition, low-latency internal buffers are required between NTT stages to support data reordering.
These features align well with field-programmable gate arrays (FPGAs), which offer fine-grained parallelism through custom cores and highly configurable memory architectures.

%
A number of FPGA-based NTT accelerators have been presented, targeting not only PQC schemes but also homomorphic encryption schemes \cite{yang2022nttgen, lu2024ntt, choi2026streamntt, ye2021fpga, wang2023sam, riazi2020heax, kim2020hardware}.
For example, NTTGen is a framework that generates optimized, low-latency NTT designs through manual register-transfer level (RTL) coding~\cite{yang2022nttgen}.
In contrast, AutoNTT~\cite{kumarathunga2025autontt} and StreamNTT~\cite{choi2026streamntt} adopt a high-level synthesis (HLS)-based approach, which translates C/C++ or OpenCL code into hardware description language (HDL) code and enables rapid architectural exploration and parameterization.
Specifically, StreamNTT is well-suited for high-throughput PQC schemes because it maximizes parallelism by deploying hundreds to thousands of butterfly units (BUs). 
These units are arranged and executed in parallel within a stage, across stages, and even across multiple NTT instances.

%
StreamNTT is released as open-source software to enhance transparency and encourage collaborative innovation\footnote{\url{https://github.com/applesforme/StreamNTT}}.
However, it relies heavily on commercial toolchains and vendor-specific FPGA platforms. 
Therefore, we present a reproducible StreamNTT benchmark built on the Verilog-to-Routing (VTR) toolchain~\cite{elgammal2025vtr}, enabling fully open, end-to-end evaluation from RTL through place-and-route.

\section{StreamNTT}
%
%
StreamNTT \cite{choi2026streamntt} uses several optimization techniques.
The first optimization is about the HLS coding style.
In NTT, the stride changes between stages, so a reorder buffer is typically used to adjust the element order between stages.
Earlier HLS-based NTT designs often implement buffer read, butterfly computation, and buffer write as separate pipelined loops, which creates three distinct dataflow modules~\cite{nguyen2019high}.
In Vitis HLS, communication between dataflow modules is limited, and pipelining occurs only within each module. 
Therefore, the three phases are executed serially rather than concurrently.
To address this issue, StreamNTT splits the reorder buffer, relocates the resulting segments into the BU dataflow module, and operates them as circular buffers.
This modified BU is called the integrated circular BU (ICBU).
StreamNTT expresses ICBU as a single pipelined loop in HLS, which enables pipelining to span across buffer read, butterfly computation, and buffer write phases.

%
The second optimization targets BU merging.
Without it, each ICBU is implemented as an independent dataflow module, leading to inefficient hardware use.
In a streaming architecture, first-in-first-outs (FIFOs) between modules increase resource overhead, and sharing precomputed twiddle-factor ROM becomes challenging.
Merging multiple ICBUs into a single dataflow module reduces these overheads.
However, excessive merging can limit parallelism.
Large modules spanning multiple FPGA super logic regions (SLRs) also complicate placement and timing, potentially degrading performance.
To address this, ICBUs are selectively merged based on stage stride, forming two groups.
The L-stage module consolidates early-stage ICBUs into a linear dataflow module, eliminating internal FIFOs and sharing twiddle-factor ROM.
The X-stage module combines later-stage ICBUs, further removing internal reorder buffers.

%
The third optimization targets placement-aware instance-level parallelism. 
In hardware designs with multiple levels of parallelism, a key question is whether to implement a single large instance for high throughput or multiple smaller instances operating in parallel.
NTT faces the same dilemma.
A single large NTT instance often spans multiple SLRs, creating placement and timing challenges, which are exacerbated when connected to multiple high-bandwidth memory (HBM) channels. 
To address this, StreamNTT deploys multiple smaller, independent NTT instances, each linked to a dedicated pair of HBM channels. 
The number of BUs per instance is chosen based on HBM bandwidth. 
This strategy improves scalability, simplifies floorplanning, and enables higher clock frequencies.

\begin{figure}[t]
  \centering
  \includegraphics[width=.9\linewidth]{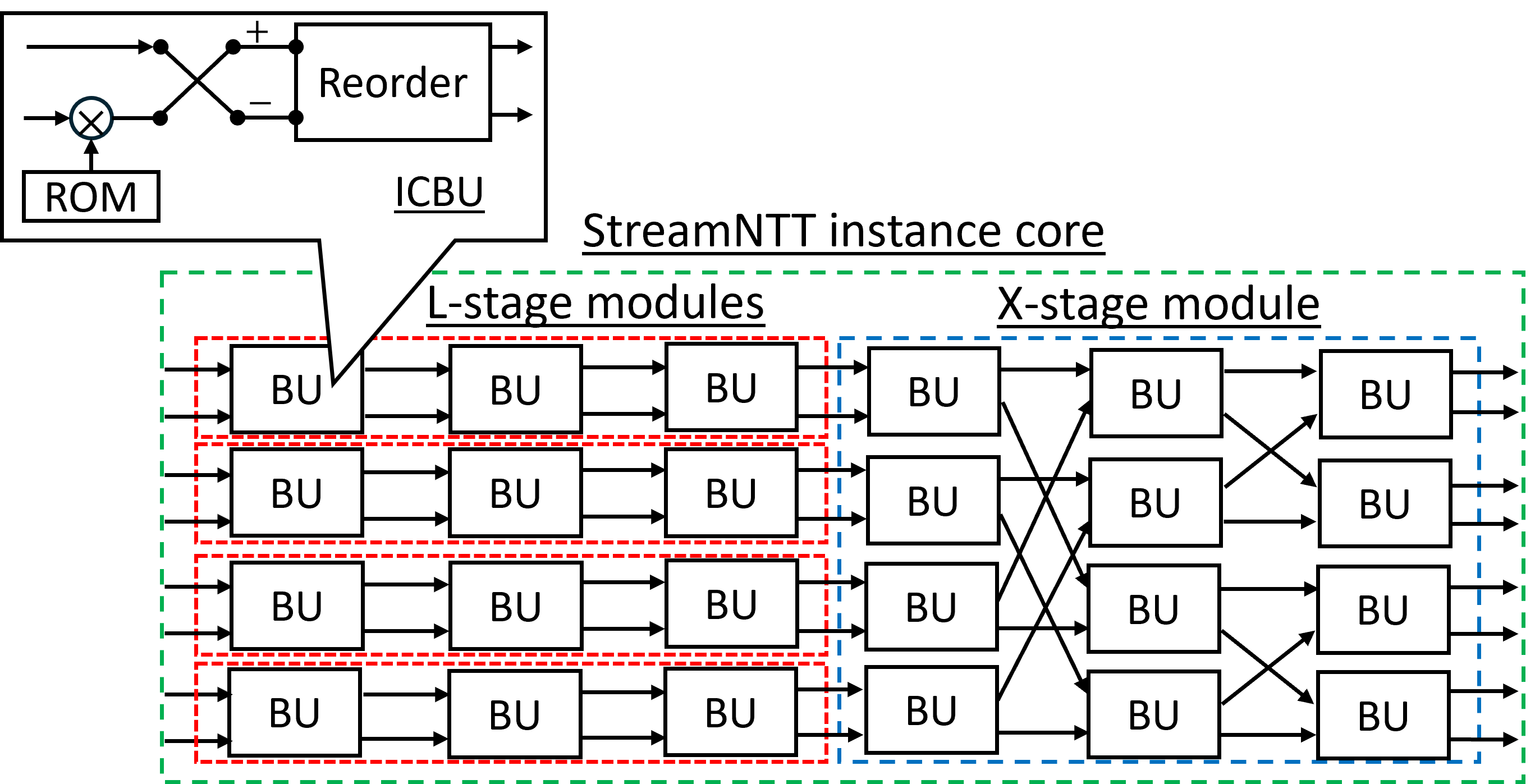}
  \vspace{-0.1mm}
  \caption{Hardware architecture of a StreamNTT instance. The ICBU performs butterfly operations, and the precomputed twiddle factors stored in on-chip ROM are shared among multiple ICBUs within the same module.}
  \vspace{-1mm}
  \label{fig:arch}
\end{figure}

Fig. \ref{fig:arch} shows the StreamNTT hardware architecture (a single NTT instance), and Table \ref{tab:origin} presents the post-place-and-route implementation results of the full StreamNTT accelerator on the AMD Alveo U280 for an NTT size of 1,024 and a modulus of 3,221,225,473.

\section{StreamNTT on Verilog-to-Routing Toolchain}
%
%
This section presents the implementation of StreamNTT in a VTR-based environment. 
Reproducing components outside the NTT core logic, such as HBM interfaces and AXI interfaces, is challenging in VTR.
Therefore, we implement and evaluate four computational kernels: a single ICBU; L-stage modules (total); an X-stage module; and a complete StreamNTT instance core.
The implementation process we used in a VTR-based environment is as follows:
\begin{itemize}
    \item 1) Verilog files are generated during the Vitis HLS synthesis process from the StreamNTT HLS code, prior to vendor-specific backend processing (e.g., FPGA synthesis and place-and-route).
    \item 2) They are then sanitized and synthesized using Yosys \cite{wolf2013yosys}, and mapped onto the VTR architecture model k6FracN10LB, which is a generic 22nm academic COFFE FPGA architecture~\cite{chiasson2013coffe}, featuring K6 (6-input) fracturable lookup tables (LUTs), 10 functional logic elements per configurable logic blocks, 20Kb memory blocks, and complex digital signal processing (DSP) blocks. 
    \item 3) A top-level wrapper instantiates each kernel module (i.e., an ICBU, L-stage modules, an X-stage module, or an NTT instance core) and exposes the necessary I/O ports. Minor modifications are applied to ensure compatibility of the HLS-generated RTL with Yosys and VTR.
\end{itemize}


\begin{table}[t]
    \centering
    \caption{FPGA Implementation Results of StreamNTT \cite{choi2026streamntt}}
    \vspace{-0.5mm}
    \label{tab:origin}
    \begin{tabular}{c c c c c c c}
        \hline
        \multirow{2}{*}{LUT} & \multirow{2}{*}{FF} & \multirow{2}{*}{DSP} & \multirow{2}{*}{BRAM} & Freq. & Thrpt. & Energy \\
        & & & & (MHz) & (MPoly/s) & (uJ/Poly) \\
        \hline
        648K & 605K & 2,560 & 536 & 306 & 32.4 & 2.1 \\
        \hline
    \end{tabular}  
\end{table}
%
%
To validate the implementation, we also develop a version that is built on a commercial toolchain. 
It targets the AMD Alveo U280 platform, using Vitis HLS (v2023.2) for high-level synthesis, TAPA (v0.1.20250803)~\cite{chi2021extending} for task-parallel dataflow compilation, and RapidStream (v2025.1.0807)~\cite{guo2023rapidstream} for floorplanning and inter-module pipelining.

%
Table \ref{tab:performance} compares the Vitis-based (Flow A) and VTR-based (Flow B) implementations.
To facilitate a direct comparison, we first compare the DSP usage in Flow A and the multiplier (Mult) count in Flow B. 
The results show a close correspondence across all design points, with differences of no more than five units.
These minor discrepancies arise from differences in how each toolchain infers and packs multiplication operations. 
Specifically, Vitis HLS targets Xilinx DSP48E2 blocks with specific packing rules, whereas VTR maps to the generic "complexDSP" block defined in the architecture model.
Despite these differences, the strong correlation (within 3-10.4\%) provides indirect evidence that the VTR-based implementation preserves the arithmetic structure of the Vitis-based implementation.

Although a direct comparison is difficult, we also examined internal memory usage.
Flow B consistently reports higher Mem block usage than the block RAM (BRAM) count of Flow A, except for the X-stage module, which reflects architectural differences between the two memory primitives. 
Specifically, 20Kb Mem blocks in VTR may have different aspect ratio constraints, requiring multiple blocks to implement wide, shallow buffers that would fit in a single Xilinx BRAM18K. 
Furthermore, circular buffers in ICBU require true dual-port access to support simultaneous read and write at different addresses.
Vitis HLS efficiently maps these buffers to the native dual-port capability of BRAMs, whereas VTR may require additional blocks or logic to emulate equivalent functionality. 
These findings suggest that internal memory mapping is a key area where academic CAD tools diverge from commercial flows, which highlights an opportunity for future optimization in open-source synthesis tools.

\begin{table}[t]
    \centering
    \caption{Comparison of Vitis (after synthesis)- and VTR (after place-and-route)-based Implementations}
    \vspace{-0.5mm}
    \label{tab:performance}
    \begin{tabular}{ l | c c | c c}
        \hline
        \multirow{2}{*}{Design kernel} & \multicolumn{2}{c|}{Flow A (Vitis)} & \multicolumn{2}{c}{Flow B (VTR)} \\
        \cline{2-5}
          & DSP & BRAM & Mult & Mem \\
        \hline
        ICBU & 4 & 2 & 4 & 10   \\
        L-stage modules (total) & 16 & 8 & 16  & 40  \\
        X-stage module & 48 & 12  & 43 & 11 \\ 
        NTT instance & 160 & 68 & 155 & 443  \\
        \hline
    \end{tabular}
    \vspace{-1.0mm}
\end{table}

\section{Conclusion}
This work presents a reproducible benchmark for StreamNTT, a high-throughput HLS-based NTT accelerator, using academic FPGA CAD tools.
By extracting kernel modules from StreamNTT for the Yosys–VTR flow, we provide quantitative comparisons against an implementation built with a commercial toolchain.
Arithmetic resource mapping (DSP/Mult) shows strong consistency between the flows, indicating that the VTR benchmark preserves the computational structure of the HLS design.
In contrast, memory resource utilization diverges significantly due to differences in memory block granularity and dual-port inference between commercial and academic tools. 
Our benchmark artifacts\footnote{\url{https://github.com/Xiaoyi-1017/StreamNTT-VTR-benchmark}}, including sanitized Verilog and VTR project files, are publicly available to support reproducible FPGA architecture research.

\begin{acks}
This material is based upon work supported by the National Science Foundation under Grant No. 2347253.
\end{acks}

\bibliographystyle{ACM-Reference-Format}
\bibliography{bib}

\end{document}